\documentclass[aps,pre,twocolumn,groupedaddress,showpacs]{revtex4}
\usepackage{amssymb}
\usepackage{amsmath}
\usepackage{cases}
\usepackage{graphicx,color}
\definecolor{r}{rgb}{1,0,0}   
\definecolor{g}{rgb}{0,1,0}   
\definecolor{b}{rgb}{0,0,1}
\DeclareMathOperator\arccosh{arcCosh}

\begin{document}


\title{Border-Crossing Model for the Diffusive Coarsening of \\ Two-Dimensional and Quasi-Two-Dimensional Wet Foams}


\author{C. D. Schimming, D. J. Durian}
\affiliation{Department of Physics and Astronomy, University of Pennsylvania, Philadelphia, PA 19104-6396, USA}


\date{\today}

\begin{abstract}
For dry foams, the transport of gas from small high-pressure bubbles to large low-pressure bubbles is dominated by diffusion across the thin soap films separating neighboring bubbles.  For wetter foams, the film areas become smaller as the Plateau borders and vertices inflate with liquid.  So-called ``border-blocking" models can explain some features of wet-foam coarsening based on the presumption that the inflated borders totally block the gas flux; however, this approximation dramatically fails in the wet/unjamming limit where the bubbles become close-packed spheres and coarsening proceeds even though there are no films.  Here, we account for the ever-present border-crossing flux by a new length scale defined by the average gradient of gas concentration inside the borders.  We compute that it is proportional to the geometric average of film and border thicknesses, and we verify this scaling by numerical solution of the diffusion equation.   We similarly consider transport across inflated vertices and surface Plateau borders in quasi-2d foams.  And we show how the $dA/dt=K_0(n-6)$ von~Neumann law is modified by the appearance of terms that depend on bubble size and shape as well as the concentration gradient length scales.  Finally, we use the modified von~Neumann law to compute the growth rate of the average bubble area, which is not constant.
\end{abstract}



\maketitle



\section{Introduction}

Aqueous foams consist of gas bubbles in soapy water, and are prototypical far-from-equilibrium disordered systems that exhibit a variety of unusual and useful properties \cite{WeaireRivier, WeaireHutzlerBook, CantatMousses}.  Even in the absence of drainage and film rupture, the average bubble size grows or ``coarsens" with time due to the diffusion of gas from smaller / higher-pressure bubbles to larger / lower-pressure bubbles \cite{GlazierWeaire92, Stavans93}.  For two-dimensional foams, von~Neumann famously pointed out that the rate of change of area for an $n$-sided bubble is
\begin{equation}
	\frac{dA}{dt}=K_0(n-6),
\label{vn}
\end{equation}
and hence depends only on topology and not at all on details of size, shape, or constellation of neighbors \cite{VonNeumann}.  Avron and Levine showed how the von~Neumann law generalizes if the foam is embedded in curved space \cite{AvronLevine92, RothPRE12}, while Mullins \cite{Mullins89}, Glazier \cite{Glazier93}, and MacPherson and Srolovitz \cite{MacPhersonSrolovitz2007} discussed how it generalizes to higher dimensions.  An important feature is that the bubble size distribution can evolve into a self-similar scaling state, where its shape is constant, and the average bubble radius grows as the square-root of time \cite{Mullins86}.  These theories all pertain to mathematically ideal dry foams, where neighboring gas bubbles are separated by soap films of zero thickness.  According to Plateau's rules for local mechanical equilibrium, films meet by threes at equal $120^\circ$ angles in a so-called Plateau border (a vertex in $d=2$ dimensions, a curve in $d=3$); and borders meet by fours at equal $\arccos(-1/3)$ angles in a vertex ($d=3$).

This paper concerns the coarsening rules for physical foams that are not mathematically dry, but have some appreciable liquid content that may be varied from nearly a dry limit up to and beyond a wet limit where the bubbles become close-packed circles or spheres.  While some of the liquid resides in the films, which have a constant thickness $l$ of order $1-100$~nm as set by disjoining pressure \cite{Israelachvili}, it mostly resides in the borders and vertices.  This follows from the typical hierarchy of length scales, where $l$ is much less than border/vertex thickness $r$, which in turn is less than the average bubble radius $R$.  Then the liquid fraction scales as $\varepsilon \sim (l R + r^2)/R^2 \sim (r/R)^2$ in $d=2$ and $\varepsilon \sim (lR^2 + r^2R + r^3)/R^3 \sim (r/R)^2$ in $d=3$; i.e.\ the Plateau border thickness is $r=\mathcal O(\sqrt{\varepsilon}R)$ and thus changes with both wetness and bubble size.  In $d=2$ there is a decoration theorem showing that borders can inflate in this manner without any change in the location of the underlying ``undecorated" dry films or vertices \cite{BoltonWeaire91, BoltonWeaire92, Weaire99, Fortes05, Mancini07}.  Though this breaks down when neighboring borders and vertices merge, and though the theorem does not hold in $d=3$, the concept of decoration provides intuition for the structure of not-very-wet foams.

The coarsening process must be slowed by wetness, since gas diffusion is faster across films than across the thicker borders and vertices.  This can be examined by the growth of the average bubble radius versus time, $R(t)$ versus $t$, where $R$ is proportional to the average bubble volume raised to the power of $1/d$.  In the dry limit for $d=2$, the expectation is $R(t)\sim t^{1/2}$, since $d\langle A\rangle/dt = 2K_0[\langle A\rangle^2/\langle A^2\rangle][\langle\!\langle n\rangle\!\rangle-6]$ follows from von~Neumann and self-similarity \cite{RothPRE13}, where $\langle\!\langle n\rangle\!\rangle \approx 6.5$ is the area-weighted average side-number and $\langle A^2\rangle/\langle A\rangle^2\approx 1.7$ \cite{RothPRE13}.  In the $\varepsilon=0$ dry foam limit, one can alternatively argue $dR^d/dt \propto D [(h \gamma/R)/l]R^{d-1}$ where $D$ is gas diffusivity, $R^{d-1}$ is the bubble surface area, and the term in brackets is the typical concentration gradient of dissolved gas in the film as set by gas solubility (Henry's constant $h$) times typical pressure difference (film tension $\gamma$ times curvature $1/R$) divided by film thickness; this also gives $R(t)\sim t^{1/2}$, but now in all dimensions.  In the $\varepsilon=1$ limit of dilute spherical bubbles, the concentration gradient at the bubble surface is instead $h (\sigma/R)/R$, where $\sigma$ is the gas-liquid surface tension; this changes the growth law to $R(t)\sim t^{1/3}$ in all dimensions.  

These arguments lead to two conflicting empirical approaches for quantifying the effect of wetness on the growth rate of the average bubble radius, $R(t)$ versus time.  One is to fit data for the growth exponent in $R(t)\sim t^\beta$, expecting $\beta$ to decrease with increasing liquid fraction.  For example $\beta=0.45$ was observed by diffuse light transmission for a foam with $\varepsilon=0.08$ \cite{DurianWeitzPine91b}; Potts model simulations gave $\beta\approx 1-\varepsilon^{1/5}/6$ in $d=2$ \cite{FortunaPRL12} and $\beta=\{0.5, 0.44, 0.4\}$ for $\varepsilon=\{0, 0.05, 0.2\}$ in $d=3$ \cite{ThomasCSA15}; and a fast crossover from $\beta=1/2$ for $\varepsilon<0.25$ to $\beta=1/3$ for $\varepsilon>0.35$ was found also by diffuse light transmission \cite{IsertEPJE}.  To our knowledge it has not been pointed out that such approaches are consistent with the assumption that the average concentration gradient around a bubble is $h (\sigma/R)/e$, set by an effective diffusive thickness $e\propto l^{1-\alpha}R^\alpha$ between bubbles across which gas must diffuse; this gives $\beta=1/(2+\alpha)$.  The second  approach assumes $\beta=1$ and analyzes data in terms of $RdR/dt=\mathcal D F(\varepsilon)$, where $\mathcal D$ has units of a diffusion constant and $F(\varepsilon)$ describes the decrease in coarsening rate with wetness.  In ``border-blocking" models, $F(\varepsilon)$ is the fraction of a typical film that is undecorated by inflated borders/vertices, and $1-F(\varepsilon)$ is the fraction that is blocked to gas diffusion \cite{BoltonWeaire91, BoltonWeaire92}.  Refs.~\cite{HutzlerWeaire00, StoneKoehlerHilgenfeldt01} take $F(\varepsilon)$ as decreasing from 1 to 0 as $\varepsilon$ varies from 0 to random-close packing.  This overestimates the rate reduction, since inflated borders cannot totally block gas diffusion.  Indeed Refs.~\cite{VeraDurian02, FeitosaDurian06, FeitosaDurian08} measured $RdR/dt$ over a wide range of sizes and liquid fractions, and found evidence for $F(\varepsilon)=1/\sqrt{\varepsilon}$.

These empirical descriptions of average behavior are not built on an explicit treatment of microstructure and the diffusion of gas across films and inflated borders / vertices, and hence cannot recover the von~Neumann law in the dry limit.  A step in this direction was made in Ref.~\cite{RothPRE13} for a quasi-2d foam of bubbles squashed between parallel plates of separation $H$.  Repeating the von~Neumann argument, supplemented by border blocking, the growth rate for an $n$-sided bubble of area $A$ was calculated to be
\begin{equation}
	\frac{dA}{dt} = K_0\left(1-\frac{2r}{H}\right)\left[ (n-6) + \frac{6Cnr}{\sqrt{3\pi A}} \right].
\label{dadtrothpre}
\end{equation}
Here $C$ is a dimensionless shape parameter (``circularity") set by the average film curvature times a power of bubble area, scaled to be one for circular bubbles:
\begin{equation}
	C = \left( \frac{1}{n}\sum_{j=1}^n\frac{1}{\mathcal R_j} \right) \sqrt{\frac{A}{\pi}}
\label{circularity}
\end{equation}
where $1/\mathcal R_j$ is the curvature of side $j$.  By this defintion, $C$ is one for circular bubbles, positive for convex bubbles, zero for polygonal bubbles, and negative for concave bubbles.  This result holds for any liquid fraction, as long as the bubble in question satisfies the decoration theorem \cite{BoltonWeaire91} such that its neighboring Plateau borders are separated by a film with non-zero length.  Experimentally, the average and standard deviation of the observed circularities were measured to be approximately $\langle C(n)\rangle = (1-n/5.73) \pm 0.25$ in the self-similar scaling state~\cite{RothPRE13}.  Note that the $2r/H$ term in Eq.~(\ref{dadtrothpre}) slows down the overall growth rate, while the $C$ term causes a violation of von~Neumann behavior; both terms vanish for $r=0$, thus recovering the usual von~Neumann law in the dry limit.  The modified growth law of Eq.~(\ref{dadtrothpre}) was shown to account for an observed slowing and violation of the von~Neumann law that increases with wetness $(r>0)$ and for smaller bubbles \cite{RothPRE13}.  Unphysically, however, it predicts the growth rate to vanish when the border radius inflates to $r=H/2$, since then the vertical extent of the film shrinks to zero and {\it all} gas transport is assumed to be blocked. 

In this paper we significantly extend the approach of Ref.~\cite{RothPRE13} by explicitly treating the diffusion of gas across inflated Plateau borders and vertices.  For this we use physical arguments as well as numerical solution of the diffusion equation.  We begin with surface Plateau borders in quasi-2d foams, which have a particularly simple geometry, before considering bulk Plateau borders and vertices.  We then use these results to modify the von~Neumann law to model the growth of individual bubbles caused by gas transport through borders and vertices, in addition to films.  Finally, we average the modified von~Neumann law over the bubble size distribution in order to obtain the rate of change of the average bubble area.

\section{Surface Plateau Borders\label{spb}}

\subsection{Expectation}

\begin{figure}
\includegraphics[width=2.5in]{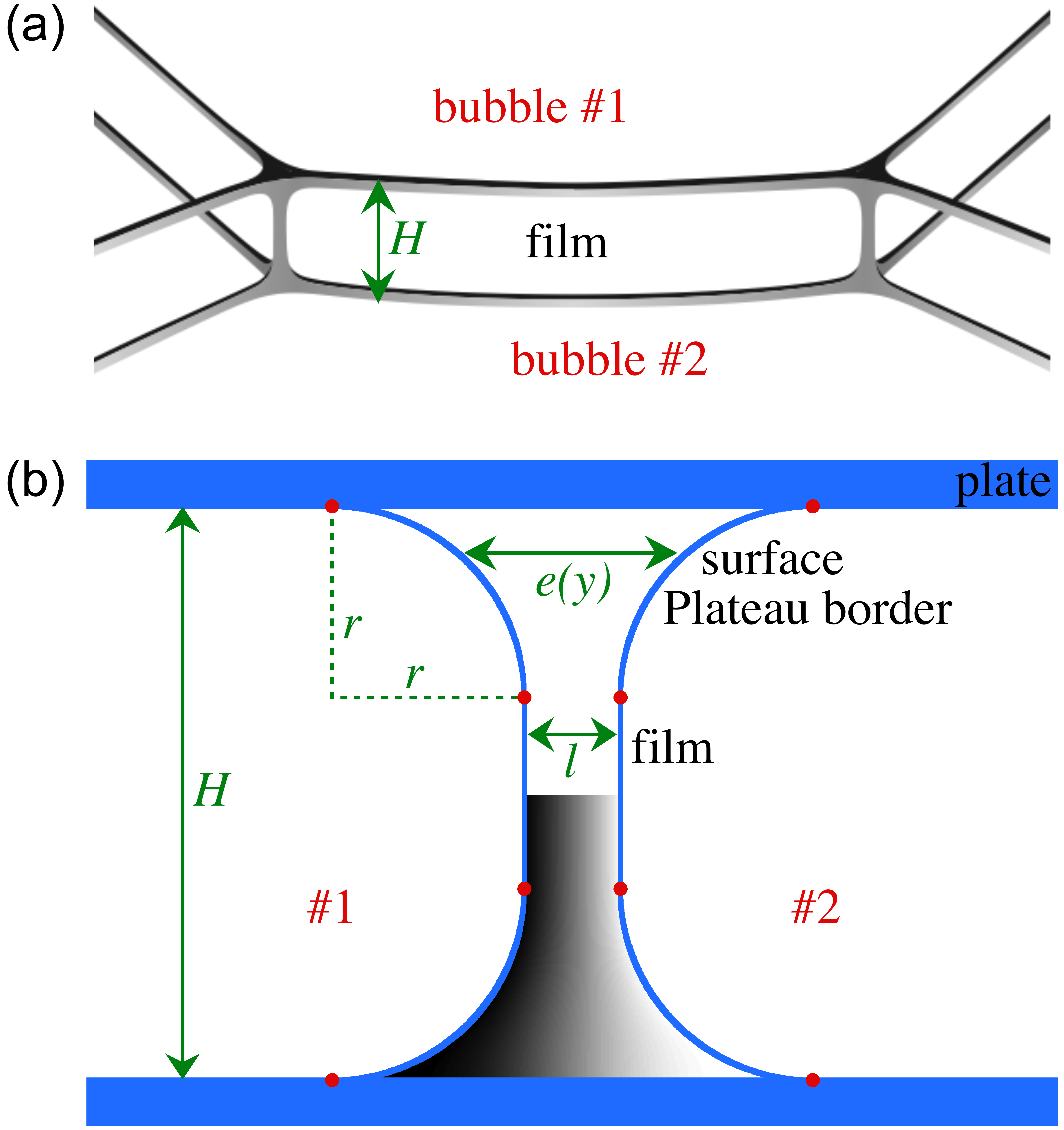}
\caption{The contact of two neighboring bubbles confined between plates of separation $H$, shown in perspective (a) and in vertical cross section (b, not to scale).  Liquid is contained in soap films of thickness $l$, in surface Plateau borders that run along the plates, in vertical Plateau borders that run between the plates, and in vertices where borders meet.  Surface borders of thickness $e(y)$ flare out from the soap film in circular arcs of radius $r$, and meet the plates tangentially; endpoints are shown as red dots.  The grayscale image in the lower half of the film/border represents the steady-state concentration field of dissolved gas, found by numerical solution of the diffusion equation, which is driven by the first bubble being under higher pressure than the second.}
\label{surfacepb}
\end{figure}

We begin by considering diffusive gas transport across Plateau borders that run along the surface of a quasi-2d foam of bubbles squashed between parallel plates of separation $H$.  Fig.~\ref{surfacepb} depicts the contact between two such bubbles, where a vertical soap film spans the gap and is surrounded by (a) two surface Plateau borders running along the plates, (b) two vertical Plateau borders running between the plates, and (c) four surface vertices where four borders meet.  A vertical cross section through the middle of the film (Fig.~\ref{surfacepb}b) shows the soap film as having thickness $l$ and terminating at distance $r$ away from the plates at surface Plateau borders of radius $r$.  In this quasi-2d geometry the liquid volume fraction scales as $\varepsilon = (RHl + Rr^2 + r^3)/(R^2H)$; therefore, the border radius inflates with liquid content as $r=\mathcal O({\varepsilon R H })^{1/2}$ assuming the usual separation of length scales.  For a sealed sample cell, $\varepsilon$ is fixed and hence $r$ grows as the foam coarsens [though more slowly than in a truly 2d sample where $r=\mathcal O(\sqrt{\varepsilon}R)$].  By contrast, for the sample cell design of Ref.~\cite{RothPRE13} it is held fixed at $r_0=\sigma/(\rho g d)$, and can be controlled by the distance $d$ of the foam above a liquid reservoir, independent of bubble size; $\sigma$ is the gas-liquid surface tension, $\rho$ is the liquid density, and $g$ is gravitational acceleration.

Two remarks on geometrical assumptions:  First, we neglect transport through the wetting layer of thickness $l_w$ on the plates, which is valid for macroscopic bubbles where $\{l, l_w\}\ll r$ holds even for very dry foams.  Second, as in the decoration theorem and the two-dimensional simulation/modeling work of Weaire and others, we assume that there is junction point separating films from Plateau borders and that the interfaces are all sections of a circle.  In reality, the detailed shape of gas-liquid interfaces in the junction region depends on the disjoining pressure and the variation of interfacial surfaces forces versus distance.  This is a good approximation for wet foams and even for rather dry foams as long as $r\gg l$ holds.

The diffusion of gas between two squashed bubbles is driven by pressure difference that leads to an imposed concentration difference $\Delta \phi$ of dissolved gas at opposite sides of the film/borders.  For two bubbles pressed into contact, at any liquid fraction, this is given by $\Delta \phi = h \gamma/\mathcal R$, the product of solubility (Henry's constant, $h$) times the Laplace pressure difference between the two bubbles (where $\gamma\approx2\sigma$ is film tension and $1/\mathcal R$ is film curvature).  The total two-dimensional diffusive current (i.e. the current per unit distance perpendicular to Fig.~\ref{surfacepb}b) may be written as the sum of contributions across the film and across the two surface Plateau borders:
\begin{eqnarray}
	I &=& I_f + 2I_{sb} \label{ifisb} \\
	  &=& D \left(\frac{\Delta \phi}{l}\right)(H-2r) + D\left( \frac{\Delta \phi}{e_2}\right)(2r) \label{l2def}
\end{eqnarray}
where $D$ is the gas diffusivity, the terms in tall braces are concentration gradients, and the geometrical factors are the vertical distances across which the gas diffuses (which sum to $H$).  While the gradient across the film is clearly $\Delta\phi/l$, since the film thickness is constant, the gradient across the surface Plateau borders is not obvious.  Therefore, Eq.~(\ref{l2def}) serves to define $e_2$ as an effective diffusive thickness that sets the size of the average gradient in the surface borders.  This term is absent, as though $e_2$ were infinite, in the border-blocking model of coarsening.

The key task is to understand the value of $e_2$.  One might guess that it is given by an average width of the Plateau border, $e_2 \propto (r+l)$.  However, we shall demonstrate that the correct value scales as $e_2 \propto \sqrt{rl}$, which is much smaller and hence implies a much greater flux of gas through the Plateau border.  We show this via rough calculation and then via numerical solution of the steady-state diffusion equation.  First, by symmetry the actual vector gradient of the concentration field at the midplane of the film-border must point parallel to the plates.  Therefore the total diffusive gas current across one border may be found by integrating over the midplane: $D(\Delta \phi/e_2) r = \int_0^r D (\nabla \phi ) dy$.  This equation is exact, but the magnitude of the gradient is unknown.  As an intuitive but uncontrolled approximation, we take it as $\nabla \phi \approx \Delta\phi/e(y)$ where $e(y) = l + 2[r-\sqrt{r^2 - y^2}]$ is the thickness of the border at vertical distance $y$ from the beginning of the film (see Fig.~\ref{surfacepb}).   Since the resulting integral is intractable, but is dominated by flux near $y=0$ where $e(y)$ is smallest, we expand as $e(y) = l + y^2/r + \mathcal O(y^4)$.  Dropping the higher order terms and evaluating the integral gives an effective thickness of
\begin{equation}
	e_2 \approx \frac{ \sqrt{rl} }{\arctan\sqrt{r/l}} \rightarrow
	\begin{cases}
		l & r \ll l~{\rm (dry)}, \\
		\frac{2}{\pi}\sqrt{rl} & r \gg l~\rm{(wet)}
	\end{cases}
\label{l2approx}
\end{equation}
In spite of approximations, both the dry and wet foam limits are actually both correct.  The former is clear; the latter will be demonstrated numerically next.  The latter also agrees with expectation based on the ``thermal conduction shape factor" that specifies the rate of heat transfer between two parallel cylinders in near contact [see e.g. Ref.~\cite{Holman} and Eq.~(\ref{Icapprox}) below].

As an aside, we point out that it is natural to use the steady-state diffusion equation since the time for a gas molecule to diffuse between bubbles is small compared to the time for bubbles to change size and to the time between rearrangements -- thanks to the usual separation of length scales.  In fact, even imposed shear does not affect the coarsening rate of a wet foam, which implies that the diffuse concentration field settles very quickly to steady state \cite{Gopal95}.

\subsection{Numerical methods}

To simulate the diffusive gas current for a given geometry, as specified by values of \{$l,r,H$\}, we solve the steady-state diffusion equation on a Cartesian grid of lattice sites $\{i,j\}$ by the method of successive relaxation \cite{NumRec, HansenLaplace}.  By symmetry, it is only necessary to simulate one quarter of the actual area.  For sites on the vertical midline between bubbles the concentration is fixed to $\phi(0,j)=1/2$, and for sites straddling the gas-liquid interface it is fixed to $\phi(i,j)=0$.  The concentration at interior sites is seeded with a profile that linearly ramps between the boundary values.  To solve $\nabla^2\phi=0$ in the interior, the concentrations at interior sites are successively updated according to
\begin{eqnarray}
	\phi_{avg} &=& (\phi_n + \phi_e + \phi_s +\phi_w)/4 \\
	\phi(i,j) &\rightarrow& (1-\omega)\phi(i,j)+ \omega \phi_{avg}
\end{eqnarray}
where subscripts $\{n,e,s,w\}$ specify sites to the north, east, south, and west of the site being updated (per notation in Ref.~\cite{HansenLaplace}); here $\omega$ is 1 for standard relaxation, or up to nearly 2 for overrelaxation \cite{NumRec, HansenLaplace}.  To enforce reflecting boundary conditions along the southern boundary of the system, $j=0$, the update rule is
\begin{eqnarray}
	\phi_{avg} &=& (2\phi_{n} + \phi_e +\phi_w)/4 \\
	\phi(i,0) &\rightarrow& (1-\omega)\phi(i,0) + \omega \phi_{avg}
\end{eqnarray}
since $\phi_s$ at a fictitious site $j=-1$ outside the system must equal $\phi_n$ in order for the vertical flux to vanish at $j=0$.  The northern boundary is treated similarly.  Here, for a given system geometry, successive updates are performed with $\omega=1.5$ until the current (next) stops changing.

The total diffusive gas current, $I$, between the two neighboring bubbles is computed by integrating the gradient of the concentration field along the midline.  As a check, the gradient is also integrated along the curved gas-liquid interface as follows.  First, the gradient in both the x- and y-directions is found at grid points closest to the curve.  Second the gradient is dotted into the unit normal, and lastly summed over boundary points with arclength increments that themselves sum to the length of the boundary.  During relaxation, the current across the midline increases while the current across the boundary decreases, and the two converge to the same value to better than 0.01\%.  This entire procedure is then repeated for several grid spacings, and the final result for the diffuse current is found by extrapolating to zero grid spacing using a linear fit.  The uncertainty in fitting parameters is typically 0.2\%, too small to display as error bars in later plots.  Lastly, the current $I_{sb}$ through one surface Plateau border is then found from the total current $I$ by rewriting Eqs.~(\ref{ifisb}-\ref{l2def}) as follows:
\begin{eqnarray}
	I_{sb} &=& \frac{1}{2}\left[ I - D\frac{\Delta\phi}{l}\left(H-2r\right) \right] \label{Isbsim} \\
	&\equiv& D\frac{\Delta\phi}{e_2}r. \label{rl2i}
\end{eqnarray}
The first expression subtracts away the contribution from the unblocked portion of the film, and the second is the definition of $e_2$.

\subsection{Expectation vs numerical results}

\begin{figure}
\includegraphics[width=3.2in]{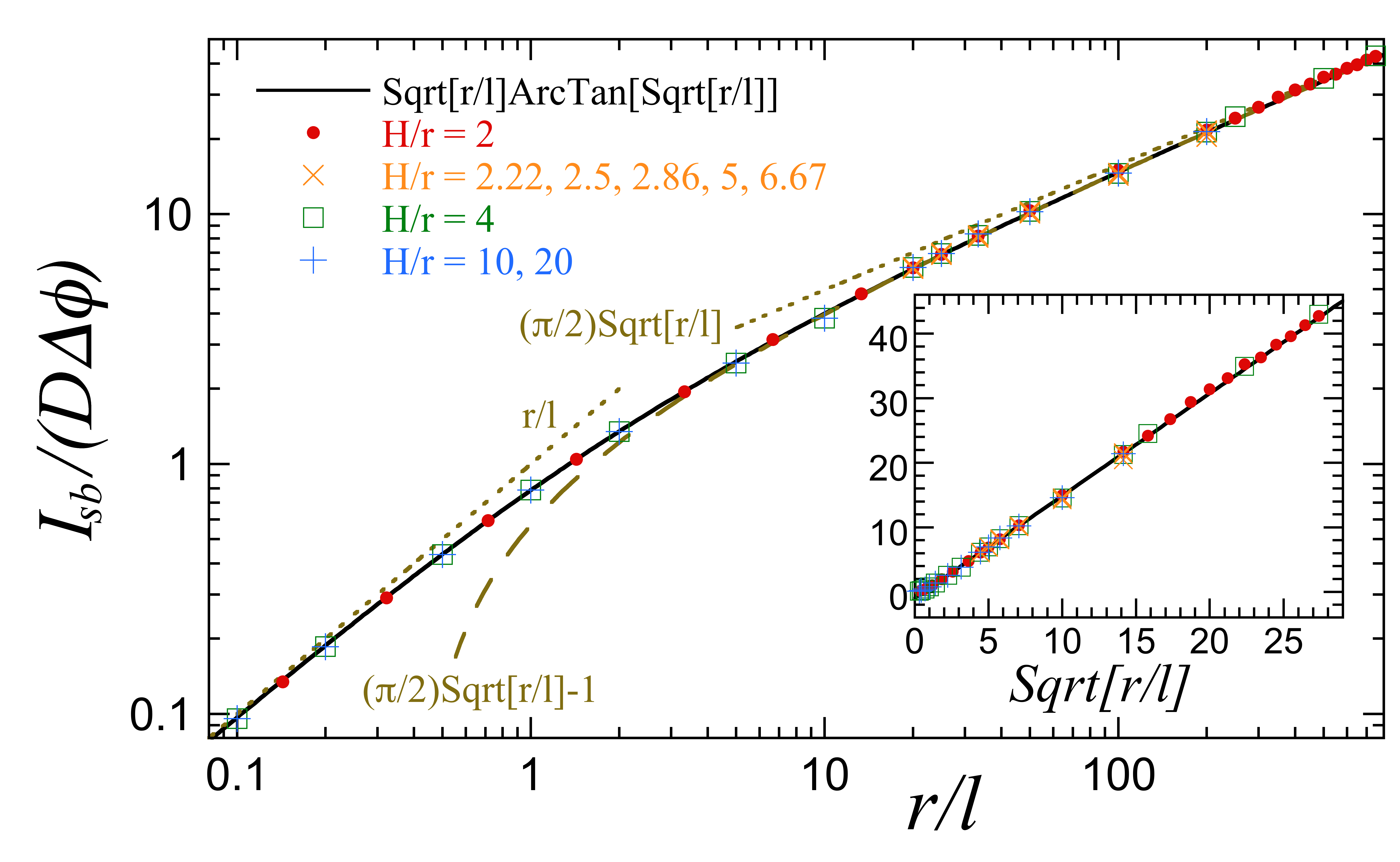}
\caption{Diffuse gas current through a surface Plateau border versus border radius $r$ divided by film thickness $l$.  Data from numerical solution of the diffusion equation are shown as symbols, for different plate separations.  The inset shows the same data along with a line fit for $\sqrt{r/l}\ge3$, given by $(1/2)(3.19\pm0.02)\sqrt{r/l}-(1.01\pm0.04)$.}
\label{sbcurrent}
\end{figure}

Simulation results for the diffusive current $I_{sb}$ through surface Plateau borders are plotted in Fig.~\ref{sbcurrent} versus the ratio $r/l$ of border curvature to film thickness, for several different values of the gap $H$ between the plates.  As $r$ increases, i.e.\ as the foam becomes more wet, the surface border current increases monotonically.  And, as expected, the data for different gaps collapse together onto a single curve.  The general behavior is seen to be $I_{sb}\propto r/l$ in the small-$r$ dry-foam limit, and $I_{sb}\propto \sqrt{r/l}$ in the large-$r$ wet-foam limit.  For comparison, the expectation from Eq.~(\ref{l2approx}) is
\begin{equation}
	\frac{I_{sb}}{D\Delta \phi} \approx  \sqrt{\frac{r}{l}} \arctan\sqrt{\frac{r}{l}} \rightarrow
	\begin{cases}
		\frac{\pi}{2}\sqrt{\frac{r}{l}} - 1 & l \ll r, \\
		\frac{r}{l} & r \ll l. \\
	\end{cases}
\label{rl2iexpectation}
\end{equation}
The full expression, and the two special limits, are all plotted along with the simulation data in Fig.~\ref{sbcurrent}.  As seen, the agreement is very good across the entire range of $r/l$ from very dry to very wet.  The large-$r/l$ wet behavior is isolated in the inset as a linear-linear plot of $I_{sb}$ versus $\sqrt{r/l}$.  Fit to a line for $\sqrt{r/l}\ge4$ gives $(1/2)(3.19\pm0.02)\sqrt{r/l}-(1.01\pm0.04)$ in close accord with the $(\pi/2)\sqrt{r/l}-1$ expectation.  The good agreement here, as well as in the main plot, demonstrate that the physics of diffusion across surface borders is now well understood.  The key feature is that the effective diffusive thickness scales as $e_2 \sim \sqrt{r l}$ for wet foams; this is a nontrivial result, and implies a much greater border-crossing current than might have been expected.

\section{Bulk Plateau Borders}

\begin{figure}
\includegraphics[width=2.5in]{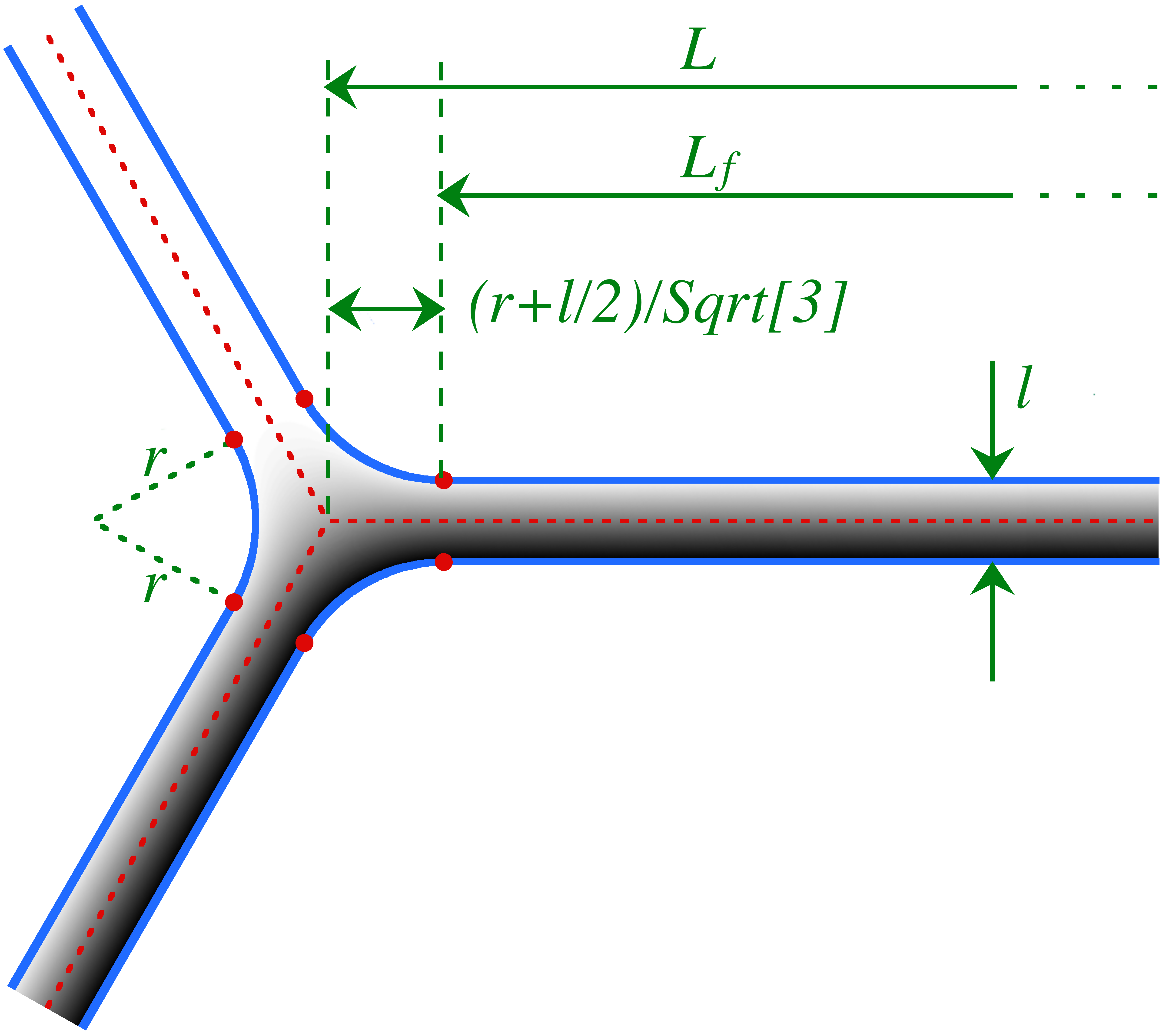}
\caption{Schematic of the gas-liquid interfaces (solid blue curves) for three neighboring bubbles, separated by films of thickness $l$ and an inflated Plateau border with radius $r$.  For this geometry, the length of the film is smaller than that in the dry limit according to $L_f=L-2(r+l/2)/\sqrt{3}$.  The grayscale image represents the steady-state concentration field of dissolved gas diffusing out of the bottom bubble.}
\label{bulkpb}
\end{figure}

We now repeat the same program for the case of bulk Plateau borders at which three bubbles meet.  This includes structures given by decoration of the point-like vertices in a truly two-dimensional foam, as well as the line-like borders in a three-dimensional foam.  As depicted in Fig.~\ref{bulkpb}, a border with radius $r$ merges with a film of thickness $l$ at distance $(r+l/2)/\sqrt{3}$ from the undecorated vertex.  Since each film flares out into part of a border at each end, the total two-dimensional current between two neighboring bubbles may be written as the sum over film plus two part-border contributions as
\begin{eqnarray}
	I &=& I_f + 2I_{b} \\
	  &=& \frac{D\Delta \phi}{l} \left(L-\frac{2r+l}{\sqrt{3}}\right)
	  + \frac{D\Delta \phi}{e_3} \left(\frac{2r+l}{\sqrt{3}}\right).  \label{l3def}
\end{eqnarray}
Note that the geometrical factors in braces sum to the undecorated film length, $L$, and that Eq.~(\ref{l3def}) serves to define $e_3$ as an effective border thickness that sets the average concentration gradient.  For comparison with simulation results for the total current $I$ between two bubbles, this can be rewritten in terms of the current through part of one of the Plateau borders:
\begin{eqnarray}
	\frac{I_{b}}{D\Delta \phi} &=& \frac{1}{2}\left[ \left(\frac{I}{D\Delta\phi}\right) - \left(\frac{L-\frac{2r+l}{\sqrt{3}}}{l}\right) \right] \label{Ibsim} \\
	&\equiv& \frac{r}{\sqrt{3} e_3}\left(1+ \frac{l}{2r}\right). \label{rl3i}
\end{eqnarray}
Just as was done for surface Plateau borders, the first expression subtracts away the contribution from the unblocked portion of the film and the second isolates $e_3$ in terms of its definition.

To predict the diffuse current through bulk Plateau borders, we again expand the thickness of the border away from the end of the film as $e(x)=l + x^2/r$.  But now we integrate to a variable distance:
\begin{eqnarray}
	\frac{I_b}{D\Delta\phi} &\approx& \int_0^{ar+bl}\frac{dx}{l + x^2/r} \\
		&=& \sqrt{\frac{r}{l}}\arctan\left( a \sqrt{\frac{r}{l}} + b \sqrt{\frac{l}{r}} \right) \label{Ibfull} \\
		&\rightarrow& \frac{\pi}{2} \sqrt{\frac{r}{l}} - \frac{1}{a} \ \ l \ll r \label{Ibapprox}
\end{eqnarray}
This is the same as the previous result for surface Plateau borders, except for the correction to leading behavior given by choice of cutoff, $a$.  Thus, for $r \gg l$, the effective diffusive thickness is expected to grow with wetness and film thickness as $e_3=e_2/\sqrt{3}\approx (2/\pi)\sqrt{r l/3}$.  Based on the sketch in Fig.~\ref{bulkpb}, the expected parameters in the full form Eq.~(\ref{Ibfull}) are $a=1/\sqrt{3}$ and $b=1/\sqrt{12}$.

\begin{figure}
\includegraphics[width=3.2in]{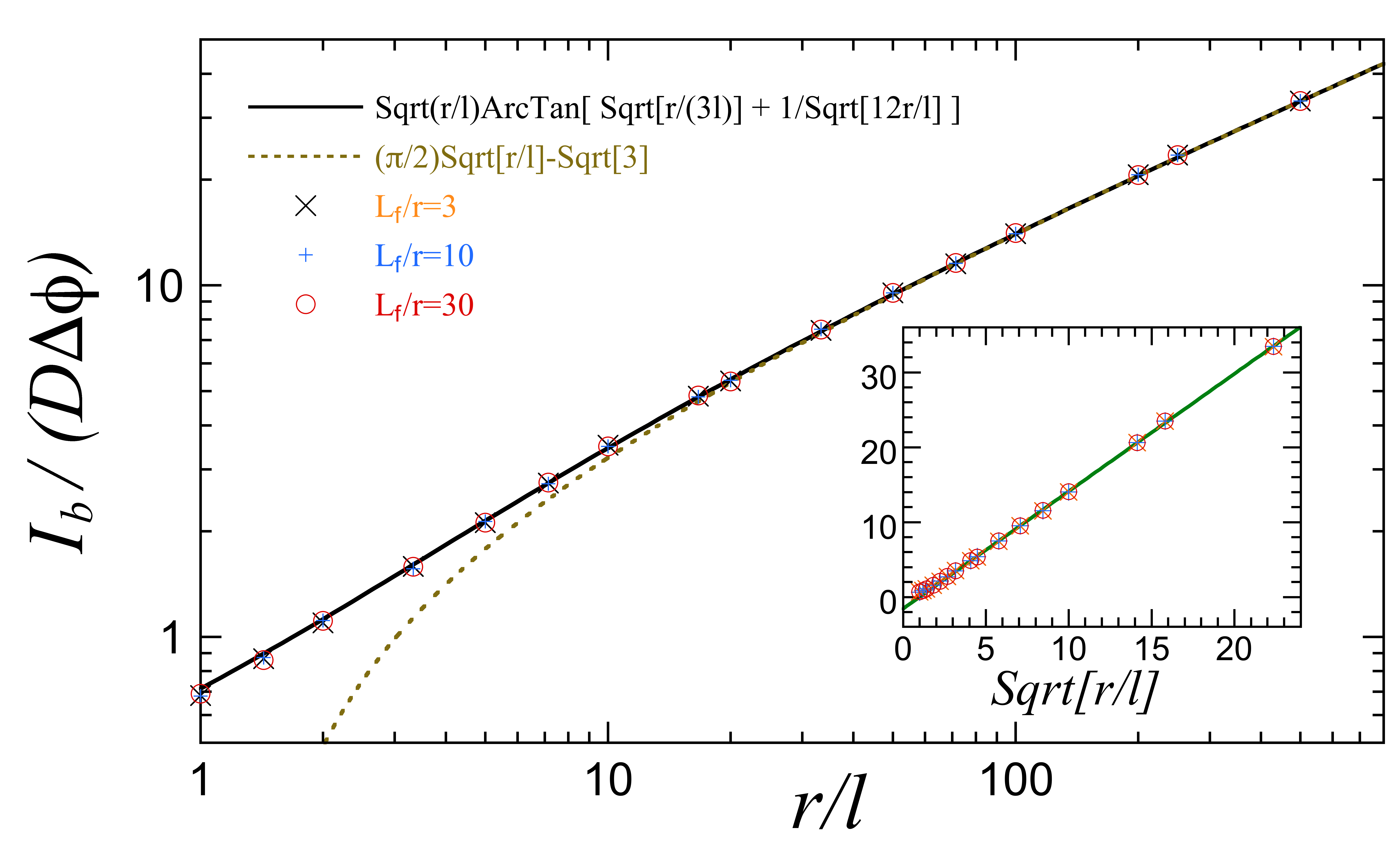}
\caption{Diffuse gas current through a bulk Plateau border versus border radius $r$ divided by film thickness $l$.  Data from numerical solution of the diffusion equation are shown as symbols, for different film lengths.  The inset shows the same data along with a line fit for $\sqrt{r/l}\ge3$, given by $y=(1/2)(3.137\pm0.001)x-\sqrt{2.43\pm0.03}$.}
\label{bulkbordercurrent}
\end{figure}

Numerical solutions of the diffusion equation for bulk Plateau borders is performed following the same procedures as for surface Plateau borders.  An example concentration field for gas leaving one of the bubbles is superposed onto the sketch of the border-film geometry in Fig.~\ref{bulkpb}.  The diffuse currents are computed across both the midline of the film, and across the gas-liquid interface, and found to agree.  Final results for the diffuse current, extrapolated to zero grid size, are plotted versus border radius $r$ divided by film thickness $l$ in Fig.~\ref{bulkbordercurrent}.  Data for three different film lengths all collapse together, showing that it's a very good approximation to separately compute and sum together the current across the film and the current across the borders.  Furthermore, the data agree very well Eqs.~(\ref{Ibfull}-\ref{Ibapprox}) with $a=1/\sqrt{3}$ and $b=1/\sqrt{12}$.  The line fit in the inset shows that a leading correction with $a=1/\sqrt{2.43\pm0.03}$ works slightly better.  Fitting the entire dataset to Eq.~(\ref{Ibfull}) gives $a=1/\sqrt{2.70\pm0.02}$ and $b=1/\sqrt{27\pm1}$.  The observed current in only slightly less than the simplest expectation, most likely because the gradient tips away from normal to the midline in moving from into the border from its junction with the film.  This effect is quite small.  We conclude that, just as for the surface Plateau borders, the physics of border crossing is well understood and is captured by the same non-trivial result that the effective diffusive border thickness scale with wetness and film thickness as $e_3\sim \sqrt{r l}$.

\section{Vertices}

In three-dimensional dry foams, Plateau borders meet at a four-fold vertex that can inflate to radius $r$ with wetness.  In quasi-2d foams of bubbles squashed between plates, three surface borders and a vertical border meet a four-fold surface vertex.  These are complicated structures, but we expect from above that transport is dominated by the region where the film begins to flare out into the vertex.  This region has area $a = \mathcal O(r^2)$ and resembles a Plateau border that is bent in the third dimension.  Therefore, the basic scaling for the three-dimensional current across both bulk and surface vertices must be $I_v\sim D(\Delta \phi/e_4)r^2$ where the characteristic concentration gradient is set by the length $e_4\sim \sqrt{r l}$.  It is beyond our scope to find the numerical prefactor or corrections to this leading behavior.

\section{Nearly-Kissing Circles and Spheres}

In very wet foams and unjammed froths, gas transport can occur between neighboring bubbles that are not actually pressed into contact.  For two-dimensional circular bubbles of radii $r_1$ and $r_2$ and center-to-center distance $r_1+r_2+l$, the diffusive current is given from the analogous problem of heat conduction between two very long parallel cylinders (see e.g. Ref.\cite{Holman}):
\begin{eqnarray}
	\frac{I_c}{D\Delta\phi} &=& \frac{2\pi}{\arccosh\left[\frac{ (r_1+r_2+l)^2-r_1^2-r_2^2}{2 r_1 r_2}\right]}, \label{Icfull} \\
	&\approx&\pi \sqrt{\frac{2 r_1r_2}{(r_1+r_2)l}}. \label{Icapprox}
\end{eqnarray}
The latter (approximate) expression holds for nearly-kissing circles, where $l$ is small compared to the bubble radii; for $r_1=r_2$, it reduces to twice the current through a surface Plateau border in the wet limit [see Eqs.~(\ref{l2def}-\ref{l2approx}): $I_{sb}\equiv D(\Delta\phi/e_2)r$ with $e_2=(2/\pi)\sqrt{rl}$].  The full expression is an overestimate for circular bubbles in a foam, because their surrounding neighbors take up some of the current.  If an effective diffusive thickness for the gap between two nearly-kissing circles is defined via $I_c = D(\Delta \phi/e_c)(2\sqrt{r_1 r_2})$, then the leading behavior from Eq.~(\ref{Icapprox}) is $e_c=(2/\pi)\sqrt{\langle r\rangle l}$ where $\langle r \rangle = (r_1+r_2)/2$.

For three-dimensional spherical bubbles, the analogous heat conduction problem is well-known only for the case that the center-to-center distance is greater than five times the larger radius (see e.g.\ Ref.~\cite{Holman}).  We are unable to find prior results for nearly-kissing spheres.  To investigate this case, we work in cylindrical coordinates where $\rho$ is the radial distance from the axis of symmetry through the spheres' centers.  To leading order, the surface-surface distance for small $\rho$ is $e(\rho) = l + \rho^2/[1/(2r_1)+1/(2r_2)]$.  Making the same approximations as the previous sections, we therefore estimate the diffuse current as
\begin{eqnarray}
	I_{s}  &\approx& D \int_0^{\rho_{m}}\frac{\Delta \phi}{e(\rho)} 2\pi \rho d\rho, \label{Iss1} \\
	&=& D\Delta\phi \frac{2\pi r_1 r_2}{r_1+r_2}\log\left[1 + \frac{\rho_m^2(r_1+r_2)}{2r_1 r_2 l}\right], \label{Iss2} \\
	&\equiv& D\frac{\Delta\phi}{e_s}\pi r_1 r_2, \label{esdef}
\end{eqnarray}
where the integration limit $\rho_{m}$ specifies the radial extent of the contact region and an effective diffusive separation $e_s$ is conveniently defined based on the area $\pi r_1r_2$.  For small $l$ and $r_1\approx r_2 \equiv r$, and taking $\rho_m=\mathcal O(r)$, the basic scaling is thus $e_s \sim r/\log(r/l)$.  Instead of scaling as $\sqrt{rl}$ per the effectively two-dimensional examples above, there is now a logarithmic factor.  

\begin{figure}
\includegraphics[width=2.0in]{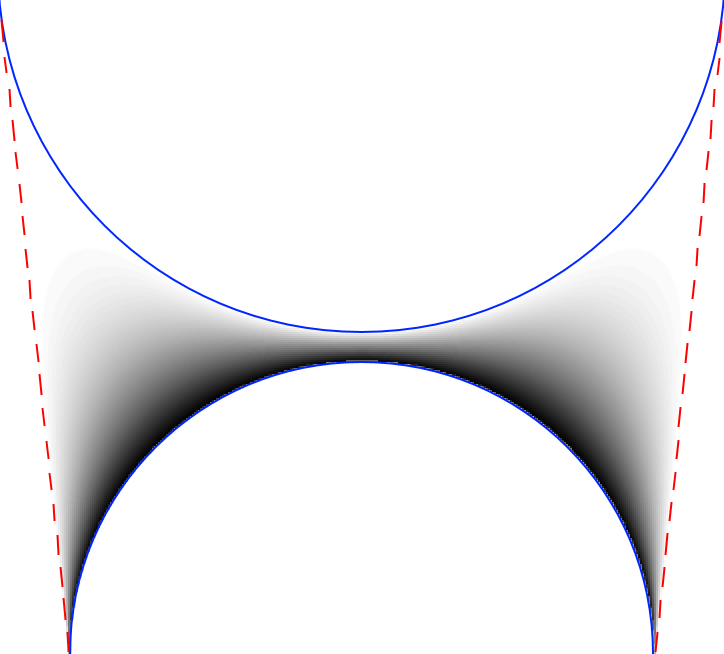}
\caption{Cross section of the diffuse gas concentration field between two spheres bounded by an absorbing cone (dashed lines).  In this example, the sphere radii are $10l$ and $12.5l$ where $l$ is the surface-surface separation along the axis of cylindrical symmetry.}
\label{twospheres}
\end{figure}

To make Eq.~(\ref{Iss2}) more concrete, we specify $\rho_m$ in terms of the length of the line segment that starts from the symmetry axis half-way between the spheres and that intersects normal to the cone that bounds the two spheres:
\begin{eqnarray}
	\rho_{m} &=& \frac{2r_1r_2+(r_1+r_2)l/2}{(r_1+r_2)+l}, \label{rhomax1} \\
	&=& \frac{2r_1r_2}{r_1+r_2} + \frac{(r_1-r_2)^2 l}{2(r_1+r_2)^2} + \mathcal O(l^2). \label{rhomax2}
\end{eqnarray}
Combining this with Eq.~(\ref{Iss2}), the leading behavior for $l \ll r_1=\mathcal O(r_2)$ is predicted to be
\begin{equation}
	\frac{I_s}{D\Delta \phi} = \frac{\pi r_1 r_2}{\langle r\rangle}\log\left[a + b \frac{r_1r_2}{\langle r \rangle l} + c\right],
\label{Iss}
\end{equation}
where $\langle r\rangle = (r_1+r_2)/2$ and $\{a,b,c\}$ are fitting parameters expected to be of order $\{1, 1, [(r_1-r_2)/(r_1+r_2)]^2\}$, respectively.

To test Eq.~(\ref{Iss}), numerical solution of the three-dimensional diffusion equation is performed in cylindrical coordinates using a two-dimensional lattice.  At $i$ lattice steps from axis of symmetry, the successive update rule then becomes
\begin{eqnarray}
	\phi_{avg} &=& \frac{\left(1+\frac{1}{2i}\right)\phi_e + \phi_n + \phi_s +\left(1-\frac{1}{2i}\right)\phi_w}{4} \\
	\phi(i,j) &\rightarrow& (1-\omega)\phi(i,j)+\omega \phi_{avg}
\end{eqnarray}
where ``north-south'' is along the symmetry axis and ``east" is in the $+i$ radial direction.  For relevance to the case of foams, where surrounding bubbles take up current, the spheres are inscribed in an absorbing cone.   An example concentration field is depicted in Fig.~\ref{twospheres}.

\begin{figure}
\includegraphics[width=3.2in]{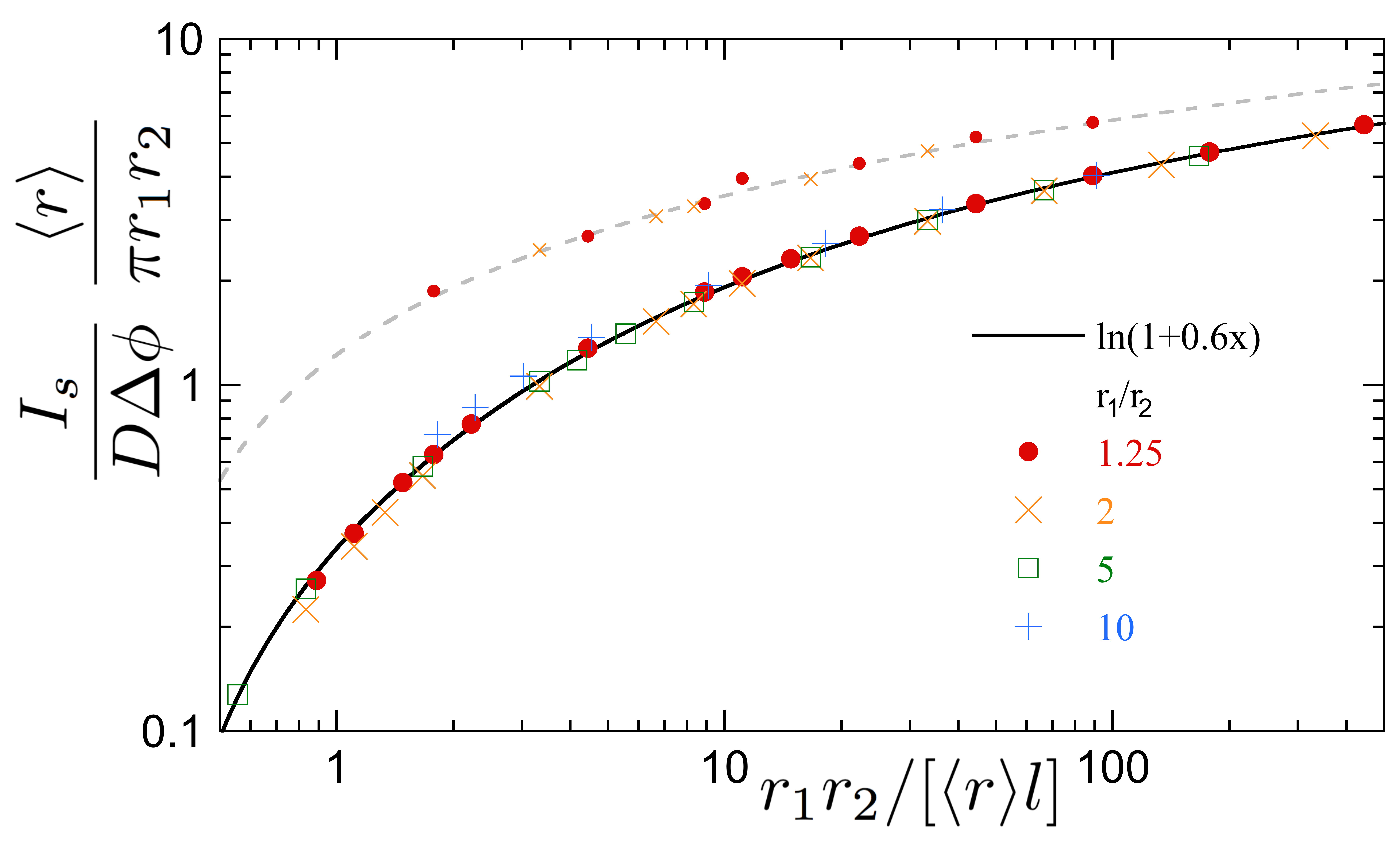}
\caption{Diffuse gas current $I_s$ between two spheres of radii $r_1$, $r_2$, $\langle r\rangle = (r_1+r_2)/2$, which are separated by a small distance $l$.  Quantities are scaled to collapse the data according to the expectation of Eq.~(\ref{Iss}).  The larger data set and solid curve are for spheres enclosed in an absorbing cone.  The smaller data set and dashed curve are for spheres enclosed in an absorbing cylinder, extrapolated to infinite radius.}
\label{twospherecurrent}
\end{figure}

The diffuse current between two spheres enclosed in an absorbing cone is computed as before, with extrapolation to zero grid spacing.  The results are scaled according to the expectation of Eq.~(\ref{Iss}) and plotted in Fig.~\ref{twospherecurrent}.  Indeed this collapses the data to the expected form, and the fit gives $a=0.8\pm0.2$ and $b=0.6\pm0.2$ with $c$ set to zero.

As an aside, data are also included in Fig.~\ref{twospherecurrent} for the diffuse current between spheres in free space.  These are obtained by enclosing the spheres more loosely in a series of absorbing cylinders, and extrapolating to infinite radius.  The fit to Eq.~(\ref{Iss}) gives $a=0\pm1$ and $b=3.4\pm0.2$ with $c$ set to zero, and is shown by a dashed curve.

\section{Modified von~Neumann Laws}

Now that the physics of border crossing is established, we explore the consequences for the growth rate of individual bubbles in wet foams.  For this,
the rate of change for the volume of an $n$-sided bubble in a quasi-2d foam can be written in terms of lengths $L_j$ and radii of curvature $\mathcal R_j$ of the undecorated films.  In particular, the sum of diffusive current across each of the $n$ sides is
\begin{widetext}
\begin{equation}
	\frac{dV}{dt} = -\sum_{j=1}^{n}\frac{D h \gamma}{\mathcal R_j}
	\left[
		\frac{(L_j-2r/\sqrt{3})(H-2r)}{l} +  \frac{ 2(L_j-2r/\sqrt{3})r}{e_2}
		+ \frac{ 2(H-2r)r/\sqrt{3}}{e_3} +  \frac{4r^2/\sqrt{3}}{e_4}
	\right]
\end{equation}
where the four terms represent contributions from the film, the two surface Plateau borders, the two vertical Plateau borders, and the four surface vertices that comprise each side.  The effective diffusive thicknesses of these elements are all $e_i=\mathcal O(\sqrt{rl})$ [recall from prior sections that $e_2=(2/\pi)\sqrt{rl}$ and $e_3=e_2/\sqrt{3}$ hold for $r\gg l$].  See Fig.~15 of Ref.~\cite{RothPRE13} for a graphic illustration of the geometrical factors in the numerators, and note that they sum to the undecorated film area $L_jH$.  We now simplify using the identity $\sum L_j/\mathcal R_j=(6-n)\pi/3$, as in the original von~Neumann argument:
\begin{equation}
	\frac{dV}{ dt} = K_0 H \left( 1 - \frac{2r}{H} + \frac{2rl}{He_2}\right)
		\left\{ (n-6) + \frac{6Cnr}{\sqrt{3\pi A}}\left[1-\frac{  \left(1-\frac{2r}{H}\right)\frac{l}{e_3} + \frac{2rl}{He_4}  }{1 - 2\frac{r}{H} + \frac{2rl}{He_2}} \right]
		\right\}
\label{dvdt}
\end{equation}
\end{widetext}
where $K_0 \equiv D h \gamma/l$ and $C$ is the dimensionless shape descriptor defined by Eq.~(\ref{circularity}).  In the dry limit of $r=0$, the bubble volume is $V=AH$ and the usual von~Neumann law is recovered, $dA/dt=K_0(n-6)$.  More generally, the factor in round braces represents an overall slowing of von~Neumann behavior, and the circularity term represents a size- and shape-dependent violation of von~Neumann behavior.  By contrast with the Ref.~\cite{RothPRE13} border-blocking version, given here in Eq.~(\ref{dadtrothpre}), the growth rate no longer vanishes in the limit of $r=H/2$, where bubble-bubble contacts are along films with infinitesimal height, since gas transport across inflated surface borders is explicitly accounted for via the new $e_2$ term.  For large bubbles and thin films, the full expression simplifies to
\begin{equation}
	\frac{dA}{dt} = K_0 \left( 1 - \frac{2r}{H} + \frac{\pi \sqrt{rl}}{H}\right)  \left[ (n-6) + \frac{6Cnr}{\sqrt{3\pi A}} \right],
\label{dadt}
\end{equation}
since contributions from the vertical borders and the surface vertices may be neglected, and both $e_2=(2/\pi)\sqrt{rl}$ and $V=AH$ hold.  For small bubbles, the relation of $dV/dt$ and $dA/dt$ is more complicated, with leading correction due to the bubble elongation (or compactness) as per Eq.~(19) of Ref.~\cite{RothPRE13}.

The generalized von~Neumann calculation (as per Eq.~\ref{dvdt}) can be repeated for bubbles in a truly two-dimensional foam, where coarsening is due to diffuse gas current across only two structural elements: the films and the three-fold vertices.  The result is
\begin{equation}
	\frac{dA}{dt} = K_0 \left[ (n-6) +  \frac{6Cnr}{\sqrt{3\pi A}}\left( 1 - \frac{\pi}{2}\sqrt{\frac{l}{r}} \right) \right]\ .
\label{dadt2d}
\end{equation}
Just as for quasi-2d foams, the usual von~Neumann law is recovered in the $r=0$ dry limit.  For $r>0$ there is no overall slowing of the growth rate, but there is a violation of von~Neumann behavior very similar to the quasi-2d case.

While our modified von~Neumann laws clearly reduce to the correct $r=0$ dry-foam limit, they cannot be applicable in the \{$r=\sqrt{A/\pi}$, $C=1$\} wet-foam limit where all bubbles are circles and the foam is at the point-J jamming/unjamming transition \cite{BoltonWeaire90, BubblePRL, BubblePRE, Epitome}.  These laws hold only for those bubbles for which each neighbor is pressed into contact across a film of non-zero length that flares out at each end into separated vertices.  With increasing wetness, the decoration theorem progressively breaks down across the sample as vertices merge and give rise to non-contacting neighbors separated by a liquid gap.  At the wet limit that the bubbles are all close-packed circles, the average number of contacting neighbors (with infinitesimal film length) must be $\langle n\rangle=4$ by isostaticity \cite{BoltonWeaire90, BubblePRL, BubblePRE, Epitome}, which is down from an average value of $\langle n\rangle = 6$ required in the dry limit by Plateau's laws and the Euler characteristic for the entire system \cite{WeaireRivier}.  For quasi-2d foams the $r=\sqrt{A/\pi}$ wet/point-J limit is distinct from the $r=H/2$ wet-but-jammed limit where bubbles are pressed into contact along films of zero height but non-zero length.

\section{Average Growth Rate}

The rate of change of average bubble area $\langle A \rangle=\sum A_i/N=A_{total}/N$ may be computed using the modified von~Neumann laws, assuming that decoration applies to each of the $N$ bubbles in the sample and that $\langle A^2\rangle/\langle A\rangle^2$ is constant (i.e. that the system is in a self-similar scaling state \cite{Mullins86}).  Given the latter, the identity $[\langle A^2\rangle / \langle A\rangle^2]\langle A\rangle = \sum A_i^2/A_{total}$ may be differentiated and rearranged to
\begin{equation}
	\frac{d\langle A\rangle}{dt} = 2\frac{\langle A\rangle^2}{\langle A^2\rangle}\sum_i^N \frac{A_i}{A_{total}}\frac{dA_i}{dt}.
\label{fullavgrowthcalc}
\end{equation}
Thus the growth rate of the average is set by the area-weighted average of the individual growth rates.  Plugging in the modified von~Neumann laws then gives
\begin{equation}
	\frac{d\langle A\rangle}{dt} = 2\frac{\langle A\rangle^2}{\langle A^2\rangle}
	 	\left\langle\!\!\!\left\langle  K\left[ (n-6) + \frac{Cnr}{\sqrt{3\pi A}} \right]
		\right\rangle\!\!\!\right\rangle  \label{fullavgrowthlaw}
\end{equation}
where $\langle\!\langle \cdots \rangle\!\rangle$ indicates area-weighted averaging, where $K$ is either $K_0$ for a 2d foam or $K_0[1-2r/H + \pi\sqrt{rl}/H]$ for a quasi-2d foam, and the near-unity factor multiplying the circularity term has been dropped assuming that the films are thin.  Obviously $\{n, C, A\}$ all vary from bubble to bubble, and this must be accounted for in the averaging.  Less obviously $r$ also varies from bubble to bubble, and hence so does $K$ for quasi-2d foams, since the pressure difference between contacting bubbles $i$-$j$ is equivalently $\gamma/\mathcal R_{ij}$ as set by film tension and curvature, or $\sigma(1/r_i - 1/r_j)$ as set by surface tension and border curvatures.

To simplify Eq.~(\ref{fullavgrowthlaw}), we consider only fairly dry foams where $r_i$ is fairly small compared to $\sqrt{A_i}$.  This is not a severe restriction, since Eq.~(\ref{fullavgrowthlaw}) already assumes that decoration holds for {\it all} bubbles in the sample (i.e.\ that there are no non-contacting neighbors).  Then the bubble-to-bubble variation of $r$ is small enough to have little effect on gas transport per unit pressure difference, and we can carry out the averages separately.  Experimental values of $\langle A^2\rangle / \langle A \rangle^2=1.72\pm0.25$ and $\langle\!\langle n\rangle\!\rangle=6.53\pm0.08$ were reported in Ref.~\cite{RothPRE13}.  The value of $\langle\!\langle Cn/\sqrt{A}\rangle\!\rangle$ has not been measured, but can be crudely estimated from tabulated data \cite{RothPRE13} for the area-weighted side number distribution $F(n)$ and the average circularities and areas of $n$-sided bubbles:  $\sqrt{\langle A\rangle}\langle\!\langle Cn/\sqrt{A}\rangle\!\rangle \approx \sum_n F(n) C_n n/\sqrt{\langle A_n\rangle/\langle A\rangle} = -0.86$ \cite{CnAhi}.  This is negative because the average is dominated by large bubbles, which have many sides and negative circularities.  Altogether, the approximate average growth rate is then
\begin{equation}
	\frac{d\langle A\rangle}{dt} \approx \alpha_1 K \left( 1 - \frac{\alpha_2 r}{\sqrt{\langle A\rangle}} \right)
\label{approxavgrowthrate}
\end{equation}
where $\alpha_1\approx 0.62$ and $\alpha_2\approx 3.2$ are wetness-independent numbers.  With increasing wetness, the average growth rate decreases due to the inflation of $r$ and also, in a quasi-2d foam, due to the resulting decrease in $K$.  For increasing dryness, the average growth rate approaches $d\langle A\rangle dt = \alpha_1 K_0$.  If the films remain at their equilibrium thickness, then this is constant; however, if the liquid pressure $p$ is lowered too far and too much liquid is sucked from the foam, then $l$ will decrease according to the stiffness of the effective interface potential, $\gamma$ will increase, and $K_0 =  Dh\gamma/l$ will increase without bound (or until the film ruptures).  In this sense, the $\{p=-\infty, r=0\}$ dry limit of a mathematically perfect ideal dry foam is physically pathological.

The solution of the growth law for $\langle A\rangle$ versus time is different for 2d and quasi-2d foams, but is asymptotically $\langle A\rangle \sim t$.  For truly 2d foams with thin films, the average border radius scales with liquid area fraction and average bubble size as $r \propto \sqrt{\varepsilon \langle A\rangle}$, as discussed in the introduction.  Therefore the average area grows at a constant rate:
\begin{equation}
	\langle A\rangle - \langle A_0\rangle = \alpha_1K_0(1-\alpha_3\sqrt{\varepsilon})t 
\end{equation}
where $\langle A_0\rangle$ is the average area at time zero (any arbitrary time as long as the system is in a self-similar state) and $\alpha_3\sqrt{\varepsilon}$ comes from the $\alpha_2$ (circularity) term in the growth law.  Note that this result assumes that (a) the film thickness and tension are constant, (b) the bubble-to-bubble variation of the border radii does not affect the gas current per unit pressure difference, (c) time zero is defined by when the foam reaches a self-similar scaling state where $\langle A^2\rangle/\langle A\rangle^2$ is constant, and (d) decoration holds, so that there are no non-contacting neighbors, throughout the entire sample.  The latter two are probably least-likely to be satisfied in principle, but all these assumptions could be tested by simulation.  

For a quasi-2d foam with these same assumptions, the growth law is different and depends on the construction of the sample cell.  For a cell where the fluid pressure is fixed by the distance $d$ of the foam above a reservoir \cite{RothPRE13}, then the Plateau border radii are nearly constant with little bubble-to-bubble variation around $r_0=\sigma/(\rho g d)$.  In this case the growth law integrates to
\begin{widetext}
\begin{equation}
	\langle A\rangle\left(1+\frac{2\alpha_2 r_0}{\sqrt{\langle A\rangle}}\right)  - \langle A_0\rangle\left(1+\frac{2\alpha_2 r_0}{\sqrt{\langle A_0\rangle}}\right) 
	                              + 2(\alpha_2r_0)^2\log\left( \frac{\sqrt{\langle A\rangle}-\alpha_2r_0}{\sqrt{\langle A_0\rangle}-\alpha_2r_0} \right)
	= \alpha_1 K_0\left(1-\frac{2r_0}{H} + \frac{\pi\sqrt{r_0l}}{H}\right) t.
\end{equation}
For a sealed sample cell with fixed liquid volume fraction, the border size is $r=\mathcal O({\varepsilon_0 \sqrt{\langle A\rangle}H})^{1/2}$, as discussed at the beginning of Section~\ref{spb}.  In this case the average growth law Eq.~(\ref{approxavgrowthrate}) is more complicated to integrate due to the variation of $K$ as well as $r$ with bubble size.  For thin films and $\varepsilon_0 \ll 1$, the result is 
\begin{equation}
	\langle A\rangle \left[ 1 + \alpha_K \left( \frac{\varepsilon_0 \sqrt{\langle A\rangle}}{H} \right)^{\frac{1}{2}}
								+ \alpha_C \left( \frac{\varepsilon_0 H}{ \sqrt{\langle A\rangle} } \right)^{\frac{1}{2}} \right] 
				- \langle A_0\rangle \left[ 1 + \alpha_K \left( \frac{\varepsilon_0 \sqrt{\langle A_0\rangle}}{H} \right)^{\frac{1}{2}} 
								+ \alpha_C \left( \frac{\varepsilon_0 H}{ \sqrt{\langle A_0\rangle} } \right)^{\frac{1}{2}} \right]
	= \alpha_1 K_0 t
\end{equation}
\end{widetext}
where $\alpha_K(\cdots)$ comes from the $2r/H$ term in $K$ and $\alpha_C(\cdots)$ arises from the $\alpha_2$ (circularity) term in the growth law.  This result is valid only if the $\varepsilon_0$ terms are small.  Thus, the average area grows nearly linearly with time for both fixed-$\varepsilon_0$ and fixed-$r_0$ sample cells, but with different characteristic features and deviations from linearity.

\section{Conclusion}

In short, we elucidated the fundamental physics of diffusive transport of gas through liquid-inflated structures in wet foams, and then we explored consequences for the coarsening process.  Namely, the diffusive current across Plateau borders and vertices is proportional to cross section times an average concentration gradient, $\Delta \phi/e$, where $e\sim\sqrt{r l}$ is a new emergent length scale for the effective thickness of the inflated structure, $r$ is the border/vertex curvature (controlled by liquid content) and $l$ is the film thickness (controlled by interfacial surface forces).  Since $e$ is much smaller than the arithmetic average thickness, $\sim(r+l)/2$, there is much more border-crossing gas transport than might have been expected.  This results in faster coarsening, as seen by new terms in the modified von~Neumann laws we derived for both 2d and quasi-2d foams.  These new terms, which cause deviation from $dA/dt=K_0(n-6)$ for individual bubbles, also cause deviation from $d\langle A\rangle/dt=$~constant, as shown in the final section.  It will thus be interesting to test for border crossing physics in measurements of the growth rates for both individual bubbles and for the average.  It will also be interesting to measure and simulate the evolution of the bubble size distribution in the transient regime, before the self-similar scaling regime is reached.


\begin{acknowledgments}
We thank Anthony Chieco and Jennifer Lukes for helpful conversations.  We dedicate this work to Dominique Langevin on the occasion of her 70th birthday.  Supported was provided by NASA grant NNX14AM99G and by a work-study grant from the University of Pennsylvania.
\end{acknowledgments}



%

%
\bibliography{FoamCoarseningReferences}

\begin{thebibliography}{37}
\bibitem[{CnA()}]{CnAhi}
\bibinfo{note}{This quantity may be rewritten in terms of the sum over all the
  films of all the bubbles: $\langle\!\langle Cn/\sqrt{A}\rangle\!\rangle =
  (1/A_{total})\sum_{i=1}^N\sum_{j=1}^n {A_i}/({\sqrt{\pi}\mathcal R_j})$. This
  is negative since the curvature of each film enters twice, once as a positive
  number (typically for smaller bubbles) and once as a negative number
  (typically for larger bubbles). A characteristic constant $\kappa$ may then
  be defined via $\langle\!\langle Cn/\sqrt{A}\rangle\!\rangle =
  -\kappa/\sqrt{\langle A\rangle}$. The crude estimate in the main text is
  $\kappa\approx 0.86$.}
\end{thebibliography}
\end{document}